\documentclass[10pt,prb,aps,twocolumn,showpacs]{revtex4}
\usepackage{graphicx,psfrag}
\usepackage{amsmath}

\def\tit#1#2#3#4#5{{#1}{\bf #2}, #3 (#4)}

\def\npb{Nucl.\ Phys.\ B\ }

\def\epjb{Eur.\ Phys.\ J.\ B\ }

\def\prl{Phys.\ Rev.\ Lett.\ }
\def\pr{Phys.\ Rev.\ }
\def\prb{Phys.\ Rev.\ B\ }

\def\jsp{J.\ Stat.\ Phys.\ }

\begin{document}

\title{Correlations and confinement in non-planar two-dimensional 
dimer models}

\author{Anders W. Sandvik}
\affiliation{Department of Physics, Boston University,
590 Commonwealth Avenue, Boston, Massachusetts 02215}

\author{R. Moessner}
\affiliation{Laboratoire de Physique Th\'eorique de l'Ecole Normale
Sup\'erieure, CNRS-UMR8549, Paris, France}

\begin{abstract}
We study classical hard-core dimer models on the square lattice with
links extending beyond nearest-neighbors. Numerically, using a
directed-loop Monte Carlo algorithm, we find that, in the presence of
longer dimers preserving the bipartite graph structure, algebraic
correlations persist. While the confinement exponent for monomers
drifts, the leading decay of dimer correlations remains $1/r^2$,
although the logarithmic peaks present in the dimer structure factor
of the nearest-neighbour model vanish.  By contrast, an arbitrarily
small fraction of next-nearest-neighbor dimers leads to the onset of
exponential dimer correlations and deconfinement.  We discuss these
results in the framework of effective theories, and provide an
approximate but accurate analytical expression for the dimer
correlations.
\end{abstract}

\date{July 12, 2005}

\pacs{74.20.Mn, 75.10.-w, 05.10.Ln, 05.50.+q}

\maketitle

\section{Introduction}

Dimer models have a venerable 
history in statistical physics.
\cite{fow37a,fis61a,kas61a,fis63a} More recently, they have also emerged as 
central models in modern theories of strongly correlated quantum
systems, e.g., high-temperature superconductors and frustrated
antiferromagnets.\cite{kiv87a,rok88a,fra91a,zengelser,milakagome,sen00a,moe01a,mis02a,sac03a,ard03a}
There, dimers can represent singlet-forming electron pairs. In order to
model the quantum fluctuations of the dimers and realize a short-range
version of Anderson's resonating valence bond (RVB) state,
\cite{and73a,and87a} Kivelson, Rokhsar, and Sethna introduced a
Hamiltonian with a resonance term which flips pairs of parallel
dimers on the two-dimensional (2D) square lattice.\cite{kiv87a} 
The purely classical dimer model also retains it relevance 
here; it was shown that the equal-weight sum over all dimer configurations 
is a ground state of the Hamiltonian when the resonance strength
equals the potential energy cost of each resonating pair of dimers
(the RK point).\cite{rok88a} The equal-time dimer correlations of the 
quantum dimer model are thus those of the classical dimers,
\cite{fn-multicrit} and the imaginary time 
dynamics can be related to a classical
random walk in configuration space.\cite{hensq,syl05a,cas05a}

The dimer pair correlations of the classical square lattice model
decay with distance as $r^{-2}$ and two inserted test monomers are
correlated with each other as $r^{-1/2}.$\cite{fis63a} 
As it turned out, away from the RK point the dimers
form long-range order and the monomers are exponentially confined.
\cite{sac89a,fra91a,leu96a} Hence this system does not give rise to
the desired RVB state with no broken lattice symmetries and deconfined
monomers (corresponding to spin-charge separation
\cite{and87a}). More recently, it was shown that a true
extended RVB phase {\it does} appear in the quantum dimer model on the
triangular lattice.\cite{moe01a} Following this insight, further work
has been carried out in order to characterize classical and quantum
dimer models on various lattices.  Moreover, there are currently
intense activities in gauge theories related to quantum dimer models.
\cite{sen00a,mis02a,ard03a,fra03a,vis03a}

To date, research on dimer models in two dimensions 
has focused mainly on planar
lattices, i.e., ones that have no intersecting links. This class of
models can be analytically solved, using Fermionic path integrals,
with the aid of a theorem by Kastelyn.\cite{kas61a} Thus the quantum
ground states at the corresponding RK points are characterized as
well. This body of analytical work has established a number of
properties of classical dimer models which are believed to hold true
even for non-planar lattices. The basis for this belief is provided by
effective theories incorporating those exact results. Such effective
theories are expected to be more robust than the exact solubility, and
therefore not crucially dependent on lattice planarity. 
Consequently, there have been relatively few numerical studies testing
these beliefs,\cite{fabsqint} although it is not necessarily the case 
that the range of behaviours unearthed so far in fact exhausts all
possibilities. In this paper, we provide a detailed study
addressing some of these issues. We do this by considering two
extensions of the square lattice dimer model which preserve its full
symmetry.

The first case we consider here
is a nonbipartite model with nearest-neighbour ($N_1$) 
and next-nearest-neighbour ($N_2$) dimers. Such bonds will inevitably
appear in realistic systems away from the limiting cases 
represented by the nearest-neighbor
($N_1$) dimer models. Introducing next-nearest-neighbor links along
only one of the diagonals destroys the square lattice symmetry and
instead yields the topology of the triangular one. It has already been
shown that an arbitrarily small fraction of such diagonal dimers
destroys the critical $N_1$ square-lattice state and leads to
deconfinement,\cite{fen02a,kra03a,ard03a} as in the isotropic triangular
lattice.\cite{moe01a}  The model with links along both diagonal
directions, i.e., the full 2D square lattice with $N_1$ and $N_2$
bonds, is not solvable by Kastelyn's theorem.\cite{kas61a} 
One would suspect (see e.g. Ref.~\onlinecite{sac00a}) 
that the critical state is 
immediately destroyed in this case as well but this has not
been explicitly demonstrated. 
Similarly, one might speculate that introducing longer bonds
between the two sublattices does not lead to deconfinement, but there
are no exact results to back this up. For this reason, we also study
the bipartite lattice with $N_1$ and fourth-nearest-neighbors ($N_4$,
of which there are eight per site). 

In both the above cases cases, we use an efficient directed-loop 
\cite{syl02a,methodnote} 
Monte Carlo algorithm to sample the full space of hard-core 
dimer configurations, with fugacities $w_i$ assigned to the different
dimer types $N_i$. We then discuss our results in the framework of
effective and simple microscopic analytic treatments. We confirm that
a low (most likely infinitesimal)\cite{heilieb} concentration of $N_2$
dimers leads to deconfinement; in particular, we find that the finite
correlation lengths for monomers and dimers induced by the presence of
the $N_2$ dimers scales inversely to their fugacity; $\xi\propto1/w_2$.  
By contrast the presence of $N_4$ dimers preserves the algebraic decay 
of dimer correlations, $\sim$$1/r^2$. These correlations have a
complex dipolar structure in reciprocal space, which implies that they
do not lead to a logarithmic divergence of the structure factor for
$w_4\neq0$; such a divergence is present for the pure $N_1$ model. For
these observations, we provide an approximate analytical theory based
on the height model in the Coloumb gas representation and a 
self-consistent mean field solution of a large-$N$ generalization
of the dimer model.

Our numerical simulations show that
the confinement exponent governing the monomer
correlation function drifts, changing continuously from $-1/2$ in
the pure $N_1$ case to $-1/9$ (to high numerical accuracy) for the
pure $N_4$ model. In the language of height models, this corresponds
to a decrease of the stiffness induced by the addition of
longer-ranged dimers. Note that for the pure $N_4$ model, like for the
soluble dimer and six-vertex models, the stiffness appears to be a
simple rational number. The efficient directed loop algorithm
provides an efficient direct determination of this quantity in 
general.

The outline of the remainder of the paper is the following: In Sec.~II we
introduce the directed-loop algorithm for a wide class of dimer models,
and in in Sec.~III we present simulation results. In Sec.~IV we 
qualitatively explain the numerical findings based on the 
height-model/Coulomb-gas representation of bipartite dimer models and 
their large-N generalizations. We conclude in Sec.~V with a summary
and a brief discussion of the relevance of our study to RVB
physics.

\section{Directed-loop algorithm}

This algorithm is an adaptation of a
quantum Monte Carlo method,\cite{syl02a,methodnote} 
with the same name, in which
updates of the system degrees of freedom are carried out along a
self-intersecting path at the endpoints of which there are defects not
allowed in the configuration space contributing to the partition
function. When the two defects meet they annihilate, the loop closes,
and a new allowed configuration is created. The conditions for
detailed balance in the process of stochastically moving one of the
defects are expressed as a coupled set of {\it directed-loop
equations},\cite{syl02a} The applicability of this scheme to dimer
models was first realized by Adams and Chandrasekharan.
\cite{ada03a} 
Here an algorithm for the multi-length dimer problem 
will be presented; a simplifying representation of the dimer configurations 
will also be introduced. The algorithm will be described only for the case 
of the $N_1$-$N_2$ model, but the scheme applies directly to any range of 
the links.

\begin{figure}
\includegraphics[width=7.5cm]{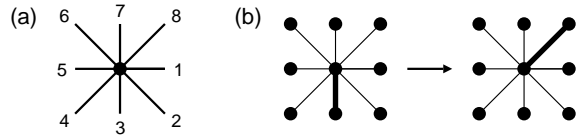}
\caption{(a) Labeling of the links of the $N_1$-$N_2$ dimer model.
(b) A step in the directed-loop update, in which the entrance to the
vertex is at site 3 and the exit is at site 8. The thick bond indicates
the location of the dimer.}
\label{fig1}
\end{figure}

The links connected to a given site are labeled as shown in Fig.~\ref{fig1}(a).
For a dimer configuration on a periodic $L\times L$ lattice, each site is 
numbered according to the type of dimer it is connected to. For any given site,
the other member of the same dimer can then be easily found. The central object
in the directed-loop algorithm is a {\it vertex}, which in this case consists 
of a site and all its surrounding sites to which it can be coupled by a dimer.
The state of the vertex is the number ($1$-$8$) assigned to the central site.
A step in the directed-loop algorithm is illustrated in Fig.~\ref{fig1}(b). 
The vertex is entered through the dimer, and one of the 
seven other surrounding sites is chosen as the exit. The dimer is then 
flipped from the entrance to the exit, the central vertex site remaining 
connected to it. The exit site already has another dimer connected to it, and 
this site now becomes the entrance in the next step of the algorithm. In each
step a defect is hence moved one link ahead, leaving behind a healed dimer 
state. The first entrance site is chosen at random and the corresponding 
defect remains stationary. It is annihilated when the moving defect reaches 
it, whence a new allowed dimer configuration has been generated. In the 
present case the defects are monomers, and the intermediate two-monomer 
configurations can thus be used to determine the correlations between two 
inserted test monomers.

The key to an efficient algorithm of this kind is that the probabilities for 
random selection of the seven possible exit sites can be chosen in such a way
that detailed balance is satisfied without any further accept/reject criterion.
Each step then moves one dimer, and a full loop can accomplish very 
significant changes to the dimer configuration. The directed-loop
equations \cite{syl02a} give the conditions for detailed balance in terms of 
weights $a_{jk}$  for the processes in which a vertex in state $j$ is entered 
at site $j$ and exited at $k$ (transforming the vertex into state $k$). 
The actual probabilities $P_{jk}=a_{jk}/w_j$, 
which implies $\sum_k a_{jk}=w_j$. Detailed balance is satisfied if $a_{jk}=
a_{kj}$. In the $N_1$-$N_2$ model, the vertices can be classified as even 
($e$) or odd ($o$) according to the numbering of Fig.~\ref{fig1}(a); there are 
then four weights: $a_{ee},a_{oo},a_{eo},a_{oe}$. In principle, one can include
``bounce'' processes where $j=k$, but in the present case they can be excluded.
Including only the seven no-bounce exits, the directed-loop equations reduce to
\begin{subequations}
\begin{eqnarray}
w_1 & = & 3a_{oo}+4a_{oe}, \\
w_2 & = & 3a_{ee}+4a_{eo}, \\
a_{eo} & = & a_{oe}.
\end{eqnarray}
\end{subequations}
This system is underdetermined and has an infinite number of postive-definite
solutions. Here the following solution will be used: For $w_1 \ge w_2$,
\begin{subequations}
\begin{eqnarray}
a_{ee} & = & a_{oe}=a_{eo}=w_2/7, \\
a_{oo} & = & (w_1 - 4w_2/7)/3,
\end{eqnarray}
\end{subequations}
while for $w_2 \ge w_1$,
\begin{subequations}
\begin{eqnarray}
a_{oo} & = & a_{oe}=a_{eo}=w_1/7, \\
a_{ee} & = & (w_2 - 4w_1/7)/3.
\end{eqnarray}
\end{subequations}
There is no guarantee that this is the best solution, but the resulting 
algorithm performs very well and allowed for studies of lattices 
with $\sim$$10^6$ dimers. The fugacity of the $N_1$-bonds is set to 
unity (except in the case of the pure $N_2$- and $N_4$-models). The program 
was tested using known results for the pure $N_1$ case \cite{fis63a} and 
by comparing with local Metropolis simulations for small lattices.

\begin{figure}
\includegraphics[width=7.5cm]{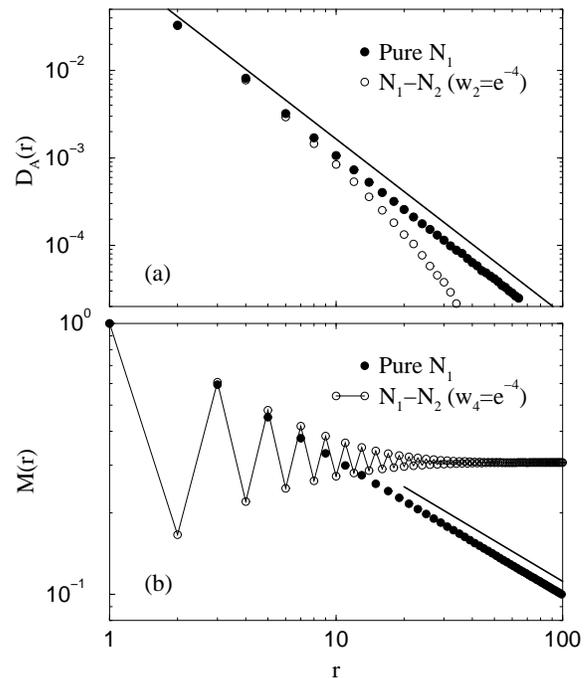}
\caption{Dimer (a) and monomer (b) correlations
along the direction $(r,0)$ for the $N_1$-$N_2$ model at $w_2=e^{-4}$, compared
with those of the pure $N_1$ model. The solid lines show the agreement with 
the known power-law \cite{fis63a} for the $N_1$ model. Note that for the pure
$N_1$ model $M(r)=0$ for even $r$. The results were obtained using $L=1024$ 
lattices.}
\label{fig2}
\end{figure}

\section{Simulation Results}

Dimer-dimer correlations in the full close-packed system and
monomer-monomer correlations in the system with two test monomers will
be discussed. The monomer correlations $M(\mathbf{r})$ were obtained
by accumulating the distances between the stationary and the moving
monomer in the directed-loop update. As has become customary, 
the normalization $M(r=1)=1$. Several types of
dimer-dimer correlations can be defined. Here $D_\Sigma(\mathbf{r})$
will be defined in the following way: If site $i$ is connected to a
dimer in the set $\Sigma$, a variable $s(i)=1$, otherwise $s(i)=0$.
The correlation function $D_\Sigma(\mathbf{r}_{ij})$ is then given by
the connected correlator $\langle s(i)s(j)\rangle$. Results will be
presented for cases where $\Sigma$ contains a single dimer or half of
the dimers of a given type (i.e., two neighboring $N_1$ or $N_2$
dimers or four neighboring $N_4$ dimers connected to a site $i$) 
For the $N_1$-$N_2$ model, the correlation
functions $D_1$, $D_2$, $D_A$, and $D_B$ are thus defined
corresponding to the sets $\{ 1\}$, $\{ 2\}$, $A=\{ 1,3\}$, and $B=\{
2,4\}$. Analogous definitions are used for correlations $D_1$, $D_A$
of $N_1$ dimers and $D_4$, $D_D$ of $N_4$ dimers in the $N_1$-$N_4$
model.

\subsection{Deconfinement in the $N_1$-$N_2$ model}

In Fig.~\ref{fig2}(a), dimer correlations $D_A$ for the nonbipartite
$N_1$-$N_2$ model with $w_2 = {\rm e^{-4}}$ (corresponding to a
concentration $p_2 \approx 0.3\%$ of $N_2$ dimers) are compared with
those of the pure $N_1$ model. There is a clear deviation from the
$r^{-2}$ decay, showing that the behaviour of the $N_1$-$N_2$ model is
fundamentally different even at this very low concentration of $N_2$
dimers. As shown in Fig.~\ref{fig2}(b), the test monomer correlation
approaches a nonzero constant, i.e., the system is deconfined in
contrast to the algebraically confined $N_1$ model. The very
significant changes seen already at a very low concentration of $N_2$
dimers suggest that an arbitrarily small concentration indeed causes
deconfinement.

\begin{figure}
\psfrag{xmonlog}{$rw_2$}
\psfrag{ymonlog}{$\sqrt{r}|D_A(r)|$}
\psfrag{w=3.0}{$w_2=e^{-3.0}$}
\psfrag{w=3.5}{$w_2=e^{-3.5}$}
\psfrag{w=}{}
\includegraphics[width=7.5cm]{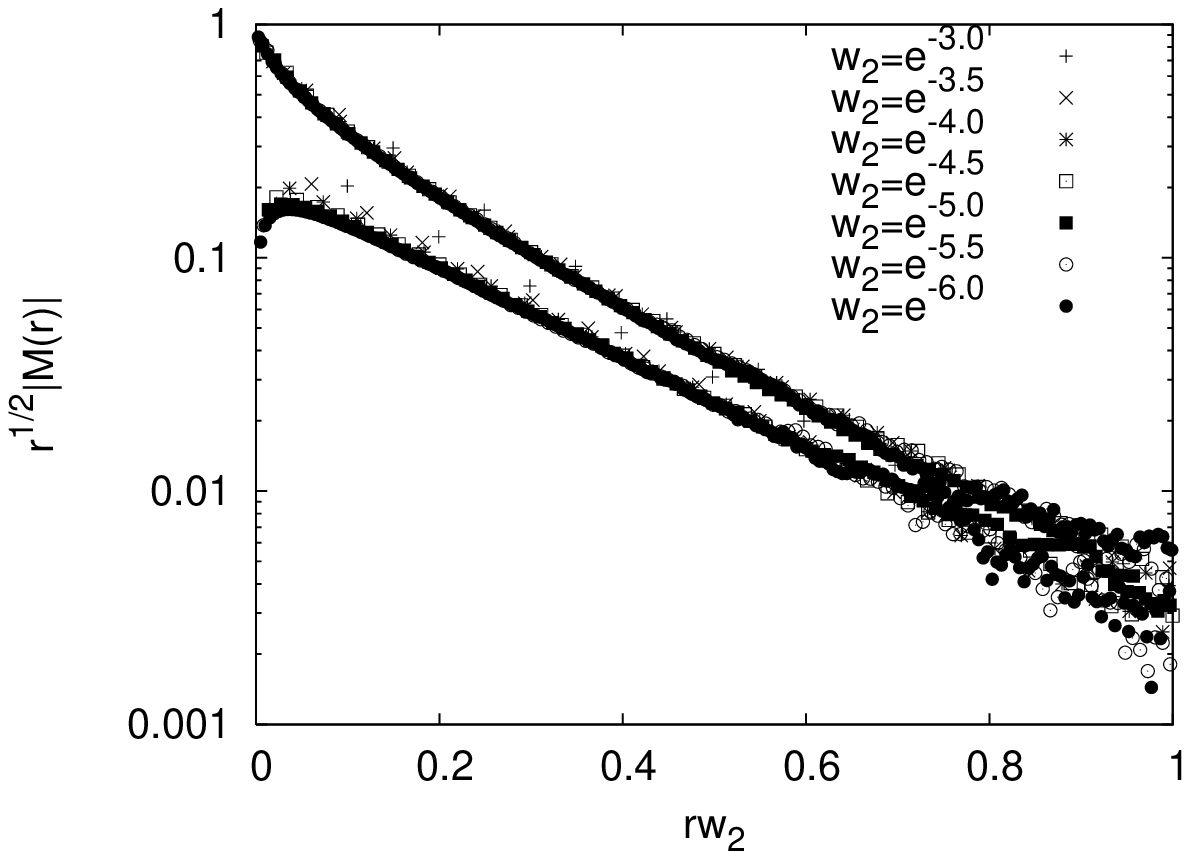}
\psfrag{xdimlog}{$rw_2$}
\psfrag{ydimlog}{$r^2M(r)$}
\includegraphics[width=7.5cm]{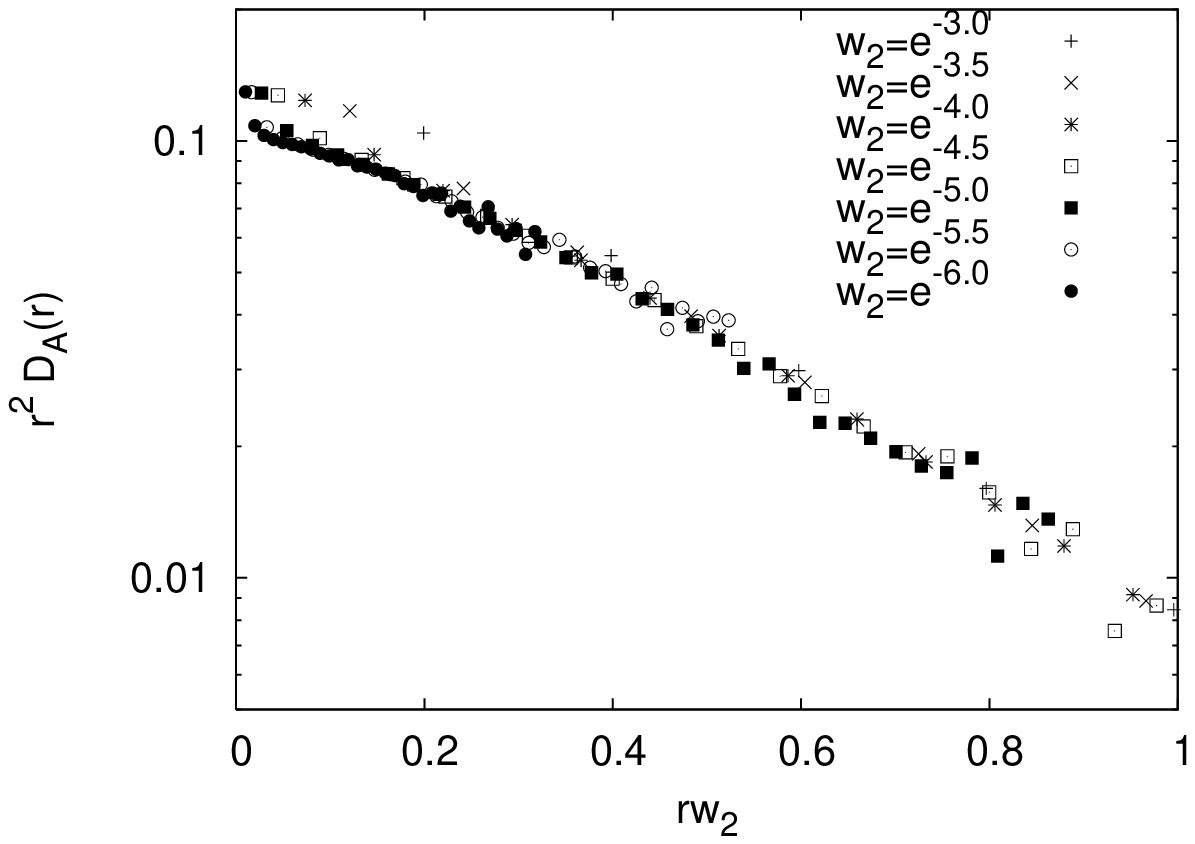}
\caption{
Scaling collapse of monomer (top panels) and dimer (bottom panels)
correlations as the fugacity $w_2$ is varied. The two distinct sets
of points in the upper panel correspond to even and odd distances.}
\label{fig:corrcoll}
\end{figure}

How does deconfinement set in upon inclusion of $N_2$ dimers?  To
address this question, we have simulated the $N_1$-$N_2$ model for
different (small) values of $w_2$ ranging from $e^{-3}\approx 1.5\%$ to
$e^{-6} \approx 0.0086\%$. We analyze the correlation functions using
a data collapse with a scaling function of the form $r^{-\alpha}\Phi(r/\xi)$. 
With $\xi\sim1/w_2$ and $\alpha$ the respective exponent for the 
correlations at $w_2=0$, we obtain a reasonable data collapse, as shown
in Fig.~\ref{fig:corrcoll}. The data for monomers is easier to handle as 
the algebraic decay is less rapid. We have also tried to improve the
data collapse by adjusting the exponent $\alpha$. We find the best-fit
values $\alpha = 2.0\pm0.1$ and $0.55\pm0.05$ for the dimers and monomers, 
respectively, consistent with them remaining at their RK point values in
this regime.

\subsection{Correlations in the $N_1$-$N_4$ model}

\begin{figure}
\includegraphics[clip,width=7.5cm]{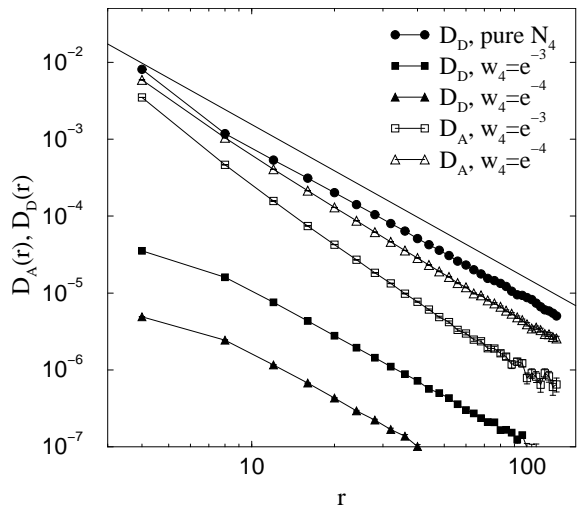}
\caption{Various dimer correlations along the direction $(r,0)$ in the 
$N_1$-$N_4$ model (on $L=1024$ lattices). The fugacities $w_4={\rm e}^{-4}$
and ${\rm e}^{-3}$ correspond to concentrations $p_4\approx 0.024$ and 
$0.065$, respectively. The solid line shows the asymptotic form $r^{-2}$.}
\label{fig3}
\end{figure}

\begin{figure}
\includegraphics[width=\columnwidth]{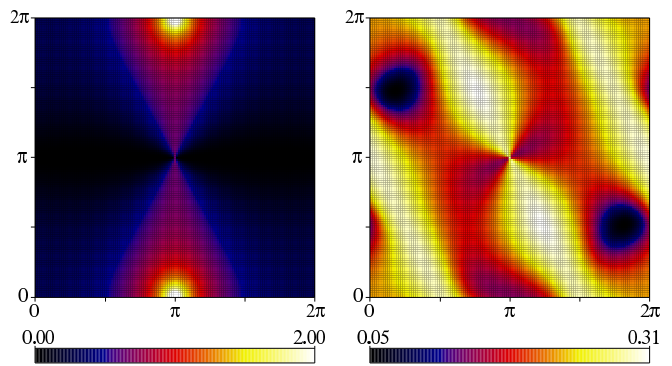}
\caption{Fourier transform of the dimer-dimer correlation $D_1$ of the $N_1$ 
model [left, for dimers pointing along $(1,0)$] and $D_4$ of the $N_4$
model [right, for dimers pointing along $(2,1$)] calculated on $L=128$
lattices. In the left graph, the log-divergent peaks at ($\pi,0$) and
($\pi,2\pi$) have been cut-off at the value $2.0$ (for this lattice
size the peak value is $\approx 3.12$)}
\label{fig4}
\end{figure}

Turning now to the bipartite $N_1$-$N_4$ model, its dimer correlations
$D_A(r)$ (involving $N_1$ dimers) and $D_D(r)$ (involving $N_4$ dimers) 
are compared with $D_A(r)$ of the
pure $N_1$ model in Fig.~\ref{fig3}. Both correlation functions appear
asymptotically to decay according to the same form $r^{-2}$ as
$D_A(r)$ of the pure $N_1$ model. Interestingly, although $D_D(r)$ is
much smaller in magnitude than $D_A(r)$ at small $w_4$, it reaches the
asymptotic form at shorter distances. In Fig.~\ref{fig4} the Fourier
transforms of the correlation functions $D_1$ and $D_4$ (involving a single 
dimer) are shown for the pure $N_1$ and $N_4$ models. For the $N_1$
model, the dominant correlations are at $\mathbf{q}=(\pi,0)$. Using
finite-size scaling, a logarithmic divergence of the peak height with 
the system size can easily be observed (not shown here).  As this feature 
masks some non-diverging correlations on the contour plot,
we have cut it off to display more clearly, most notably, a
non-analytic bow-tie feature at $(\pi,\pi)$.

In the $N_4$ model, the logarithmic peak is absent, and the
correlations display a rather more complex structure, exhibiting very
broad peaks, and again a bow-tie visible at the zone corner; this is
rotated with respect to the bow-tie for $N_1$ dimers.
These features show almost no size dependence up to the largest size
($L=2048$) studied, although one might naively have expected a
logarithmic divergence here as well, considering the $r^{-2}$
real-space correlations. 

\begin{figure}
\includegraphics[width=7.5cm]{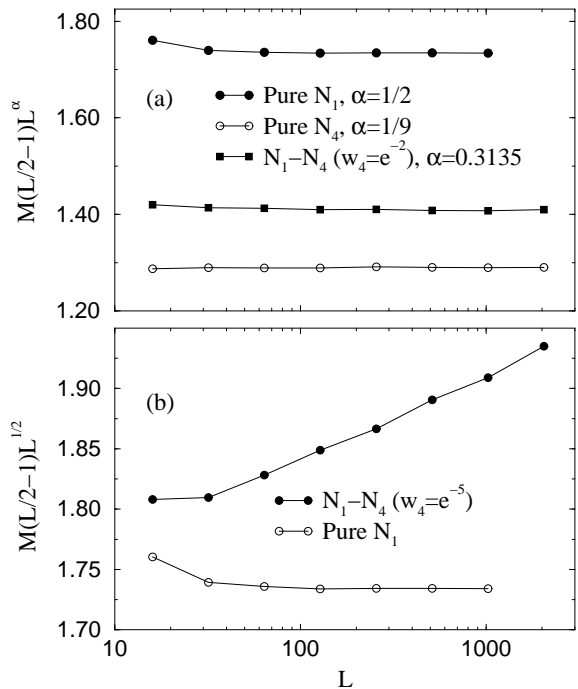}
\caption{Finite-size scaling of monomer correlations in the $N_1$-$N_4$
model. (a) The best scaling for three different cases. (b) Failure of a 
scaling with the pure $N_1$ exponent $\alpha=1/2$ for the $N_1$-$N_4$ model 
at low fugacity $w_4$.}
\label{fig5}
\end{figure}

Whereas the dimer correlations decay with a leading $r^{-2}$
dependence, a drift in the exponent of the algebraic decay is manifest
in the monomer correlations.  Results for $M(r)$ at $r=L/2-1$ are
shown multiplied by $L^\alpha$ in Fig.~\ref{fig5}(a). Here $\alpha$ is
adjusted to give a flat $L$ dependence, and hence the critical form
$M(r)\sim r^{-\alpha}$ is extracted. The known $\alpha=1/2$ is used
for the pure $N_1$ model. For the pure $N_4$ model the exponent is
consistent with $\alpha =1/9$ (to an accuracy of $1\%$). To show that
the exponent changes from $1/2$ already at a low concentration of
$N_4$ bonds, results for $w_4={\rm e}^{-5}$ ($p_4\approx 0.9\%$) are
scaled with $\alpha =1/2$ in Fig.~\ref{fig5}(b). This scaling clearly
fails, and is instead consistent with $\alpha \approx 0.485$.

\section{Analytical results}

We now present an analytical treatment of the numerical findings.  We
will in turn discuss exponents, correlation lengths, amplitudes and
the angular form of the dimer correlations.  Our discussion is based
on the height model/Coulomb gas representation of bipartite dimer
models on one hand,\cite{dombreview,zenghenley,jesjan} and on their
large-$N$ generalisation on the other.\cite{lNpap} The former is based
on a mapping of the dimer configurations onto a fluctuating height
surface and accounts for the gross features of the
correlations described above. The latter reproduces in detail the
features of the correlations of the $N_4$ model.

Let us start with a brief formulation of the height model for the
bipartite $N_1$-$N_4$ dimer model. This is analogous to Henley's work
on the pure $N_1$ model.\cite{hensq} Readers interested in deeper
detail are referred to 
Refs.~\onlinecite{dombreview,zenghenley,jesjan}.

Crucially, bipartitness allows us to orient each dimer so as to point
from one sublattice to the other. We can then assign to a dimer $n$ units
of a fictitious magnetic flux, $\vec{B}$, and to a link not occupied
by a dimer $m$ units of flux. The lattice divergence
$\vec{\nabla}\cdot\vec{B}=0$ if we choose $n+11m=0$, as in the
$N_1$-$N_4$ model, 11 of the 12 links emanate from a site are empty. It
is convenient to choose the overall scale of $\vec{B}$ so that
$n=-11m=11/12$.
The constraint $\vec{\nabla}\cdot\vec{B}=0$ can be resolved in terms
of a scalar height function, $h$, via $\vec{B}=\vec{\nabla}h$. (In
$d=2$, curl and grad are basically interchangeable). The crucial step
in the height model ansatz is to posit that, upon coarse-graining, the
entropy of the height surface gives rise to an energy functional
containing a leading quadratic term of the form
\begin{equation}
 E=\pi K \int d^2\vec{r} |\vec\nabla h|^2\ .
\end{equation}
Here, the (as yet undetermined) coupling constant $K$ is 
known as the stiffness $K$. 

Implicit in the derivation of such an energy functional of entropic
origin is a coarse-graining procedure, which maps many microstates
of the dimer model onto fewer macrostates.\cite{zenghenley}
In the process, other terms are generated both for the energy and in
the identification of operators linking the height variable to, say,
dimers and monomers. These terms are classified by an `electromagnetic'
charge, which, together with $K$, determines the exponent of the
leading power of its correlations.\cite{dombreview,zenghenley,jesjan}

First, let us discuss the $N_1$-$N_2$ model. $N_2$ dimers 
violate bipartiteness and thus 
show up as
defects in the height surface; a finite density of such defects
destroys the algebraic correlations. This was calculated analytically
for a planar dimer model interpolating between the square and the
triangular lattices by switching on only one of the two diagonal
directions of next-nearest neighbour bonds. There, a dimer correlation
length proportional to the inverse of the small fugacity was also
found, which result was supplemented by an effective field theory
arguing the same should hold true for monomer correlations in this
system.\cite{fen02a} Our numerical results, in particular the scaling
plots, confirm that these results hold also for the case of the
nonplanar $N_1$-$N_2$ model.

Next, we consider the $N_1$-$N_4$ model.  We first note that the
monomer correlations have a
power-law decay with an exponent given by $K$. The decrease of $K$
from $1/2$ upon adding $N_4$ dimers, implies that the corresponding
height model grows floppier (the heights fluctuate more strongly). We
note in passing that it it is rather straightforward to extract the
monomer correlations using the directed-loop algorithm, as this
only requires keeping track of the monomer separations in intermediate
states sampled. This thus provides a means of determining the
stiffness constant $K$ with relatively little effort.

The analytical calculation of the dimer correlations are a little more 
involved than the monomers, as they contain two
terms. One of them, the `vertex term', decays with an exponent of
$1/K$. The other, the `dipolar term', always decays as $1/r^2$; which
of these dominates depends on $K$, and for the pure $N_1$ model, they
happen both to decay with the same power law.
It is the vertex term which generates the logarithmic peak in the
$D_1$ dimer structure factor in Fig.~\ref{fig4}. Upon inclusion of
$N_4$ dimers, i.e. with increasing $K$, the power of the decay becomes
more rapid, and the divergence of this peak thus disappears when $N_4$
dimers are added. 

The survival of the leading $1/r^2$ power in the dimer correlations is
solely due to the dipolar term, which arises via the original
identification of the dimers and $\vec\nabla h$
through the lattice flux $\hat{B}$: If we choose the vectors
$\hat{e}_k$ $(k=1\cdots 12)$ to point along the direction of a dimer
from one sublattice to the other, we find a contribution to the dimer
density $n_k$ which is proportional to
$\hat{e}_k\cdot\vec{B}=\hat{e}_k\cdot
\vec\nabla h$. As long as the height correlations grow
logarithmically with distance, $\langle (h(r)-h(0))^2\rangle\propto
\ln r$, as they do for the parameters our simulations probe, the
resulting contribution to the dimer correlations always decay as
$1/r^2$, on account of the derivative linking $h$ and $\vec{B}$.
As the power $1/K$ of the vertex term is only
slightly larger than 2 
for small values of $w_4$, this leads to the very slow
approach to the asymptotic $1/r^2$ behaviour, as evidenced in our 
simulations.

Despite their $1/r^2$ behaviour, the dipolar correlations do not yield
a logarithimic peak in the structure factor. 
This is because their dipolar structure makes the
leading term vanish as the angular part of the Fourier integral
$\tilde D_A(\vec{q})=\int dr\ d\phi
\exp(i \vec{q}\cdot\vec{r})\ D_A(r,\phi)$ is performed.
Instead, what one finds are the characteristic bow-tie
structures\cite{lNpap,henbow,hermelebow} displayed in Fig.~\ref{fig4}.
These bow-ties are points of non-analyticity; there is no divergence,
but a discontinuity there. The bow-ties reflect the transverse
correlations of a divergence-free field $\vec{B}$. These bow-ties are
thus oriented with one axis along the dimer under consideration.

It has recently been observed\cite{lNpap} 
that a semi-quantitative theory for such
correlations can be obtained in the framework of a large-$N$
generalisation of the hardcore dimer model, via an intermediate
mapping to an Ising model, the ground-states of which are in
correspondance with the dimer states. This Ising model is then solved
in a self-consistent mean-field theory in the limit of low
temperatures. We note that such a generalisation does not preserve the
charge-assignments for the operators generated upon coarse-graining,
and does therefore miss the vertex term and hence the logarithmic peak
for the $N_1$ model. However, it does provide the angular dependence
of the dipolar part of the correlations.

Here, we present the relevant theory for the dimer correlations both
in the pure $N_1$ and the pure $N_4$ models. The former is of course
also soluble in closed form.  In either case, one assigns an Ising
spin $S$ to each {\em bond} of the lattice which encodes the presence
($S=+1$) or absence ($S=-1$) of a dimer on this bond.\cite{lNpap} The
hardcore corresponds to a non-zero magnetisation in the model, but
this leaves the shape of 
its $q\neq0$ correlations unchanged.\cite{henbow,lNpap}

In the case of the $N_1$ model, there are two bonds per unit cell (the
links of the square lattice in the $x$ and $y$ directions), denoted by
vectors $\vec{v}_a$, $\vec{v}_{1,2}=\hat{x},\hat{y}$. 
The relevant interaction matrix,
$J$, is thence a $2\times2$ matrix. Its Fourier transform is given by
$\tilde J_{ab}=\cos{(q_a/2)}\cos{(q_b/2)}$. Here,
$q_a=\vec{q}\cdot\vec{v}_a$, so that\cite{fn-canalsgaranin} 
\begin{equation}
\tilde{D}_A(\vec{q})\sim
\frac{\cos^2{(q_y/2)}}{\cos^2{(q_x/2)}+\cos^2{(q_y/2)}} .
\end{equation}

For the $N_4$ model, due to the non-planarity of the lattice on which
the dimer model is defined, this theory takes a somewhat more
complicated
form. One now obtains a $4\times4$ interaction matrix. Again, let
$\vec{v}_a$ denote the four different dimer directions: 
$\vec{v}_{1,2}=(2,\pm 1),\
\vec{v}_{3,4}=(\pm 1,2)$. 
The same form of the interaction matrix as above then
holds: $\tilde J_{ab}=\cos{(q_a/2)}\cos{(q_b/2)}$. We thus find the
following correlations:
\begin{equation}
\tilde{D}_4({\vec{q}})\sim
%\left[ 
1-
\frac{\cos^2{(\vec{q}\cdot{\vec{v}_1/2})}}
{\sum_{a=1}^4\cos^2(\vec{q}\cdot{\vec{v}_a/2})}
%\right]
.
\end{equation}

\begin{figure}
\includegraphics[width=\columnwidth]{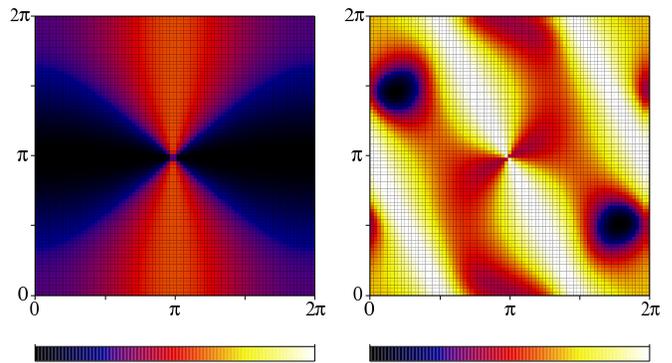}
\caption{Fourier transform of the dipolar part 
(i.e., without the dominant logarithmic peaks in $N_1$) of the
dimer correlations $D_1$ of the $N_1$ model (left) and $D_4$ of
the $N_4$ model (right), calculated from a large-$N$ theory. The
over-all amplitudes are not given accurately by the theory and
are therefore not indicated here.}
\label{fig4a}
\end{figure}

These low-temperature spin (and hence approximate dimer) correlations 
are plotted in Fig.~\ref{fig4a} above. We note that not only are the
bow-ties of the simulation data, Fig.~\ref{fig4}, reproduced in their 
correct orientation, but the qualitative structure of $\tilde D_4$ in 
reciprocal space is reproduced throughout. 

\section{Conclusions}

The results obtained here demonstrate that the 2D square-lattice dimer
model becomes deconfined when any finite density 
%a very low concentration (likely infinitesimal) 
of non-bipartite (next-nearest-neighbor) dimers are
introduced. The system remains algebraically confined in the presence
of longer bipartite (fourth-nearest-neighbor) dimers, but the
corresponding monomer exponent drifts. In contrast, as the
stiffness of the corresponding height surface decreases, the leading
dimer-dimer correlation exponent remains unchanged at $1/r^2$, due to
the persistance of the dipolar term due to the gauge structure
captured by the height model. Our results thus fit in with the known
beliefs on dimer models and can be seen as a detailed check of their 
validity. We have shown that the directed-loop simulation algorithm can
produce very accurate results for the drifting monomer 
exponent. In particular, the accuracy is high enough to suggest the 
conjecture that this exponent for the pure $N_4$
model takes the rational value of  $1/9$.

Our findings are relevant to quantum dimer models as well. Various
resonance terms can be introduced, and corresponding RK points can
then be demonstrated in the same way as has been done for other models.
\cite{rok88a,ard03a,hen03a} Hence, it is clear that an RVB state can
in principle
be realized on the square lattice once next-nearest-neighbor dimers
are allowed.  In the context of the dimer models for the Mott
instulators relevant to the high-$T_c$ cuprates, two points however
need to be borne in mind. Firstly, the Rokhsar-Kivelson quantum dimer
models allows a simple classical calculation at the RK point only if
it is possible to choose the sign of the off-diagonal matrix elements
to be negative, which property needs to be established on a
case-by-case basis for non-bipartite models (it does hold for the
bipartite case).\cite{kumarklein} In addition, the description of the 
singlet space using a dimer basis becomes overcomplete as the 
coordination of the dimer lattice increases.  At any rate, our results 
again underscore the special importance played by non-bipartitness in 
the realisation of RVB liquids in $d=2+1$.

\acknowledgments

We thank E. Fradkin, C. Henley, and S. Sachdev for discussions and comments 
on the manuscript. RM is grateful to P. Fendley, S. Isakov,
W. Krauth, K. Gregor, K. Raman and S. L. Sondhi for collaboration on
related work. In the initial stages of this work AWS was supported by
the Academy of Finland, Project No.~26175. RM is supported in part
by the Minist\`ere de la Recherche et des Nouvelles Technologies with 
an ACI grant. This research was supported in part by the National Science
Foundation under Grant No. PHY99-07949 at the KITP in Santa barbara. 
Some of the simulations were carried out on the Condor system at the 
University of Wisconsin-Madison.


\begin{thebibliography}{00}
\bibitem{fn-multicrit}
This equivalence does not extend to monomer correlations: monomers at
the RK point are deconfined whereas the classical models have confined
monomers. In fact, the RK point is `deconfined multicritical',
sandwiched between two confining crystalline 
phases.\cite{moe01a,fra03a,vis03a}
\end{thebibliography}
\end{document}